\RequirePackage{fix-cm}
 \documentclass[twocolumn]{svjour3}          
\smartqed  
\usepackage{graphicx}
\usepackage{array}
\usepackage{amsmath}
\usepackage{hyperref}
\usepackage{color}
\usepackage{makecell}
\usepackage[misc]{ifsym}
\usepackage{multirow}
\usepackage{booktabs}
\usepackage[hang]{footmisc}
\usepackage{bm}
\usepackage{algorithm} 
\usepackage{algorithmicx} 
\usepackage{algpseudocode}
\usepackage{ragged2e}

\usepackage{mathptmx}      
\begin{document}

\title{CycleGAN with Dual Adversarial Loss for Bone-Conducted Speech Enhancement
} 


\author{ Qing Pan\textsuperscript{1} \and Teng Gao\textsuperscript{1} \and Jian Zhou\textsuperscript{1} \and Huabin Wang\textsuperscript{1} \and Liang Tao\textsuperscript{1} \and Hon Keung Kwan\textsuperscript{2}}



\date{Received: date / Accepted: date}

\twocolumn[
\maketitle
\begin{abstract}
  Compared with air-conducted speech, bone-conducted speech has the unique advantage of shielding background noise. Enhancement of bone-conducted speech helps to improve its quality and intelligibility. In this paper, a novel CycleGAN with dual adversarial loss (CycleGAN-DAL) is proposed for bone-conducted speech enhancement. The proposed method uses an adversarial loss and a cycle-consistent loss simultaneously to learn forward and cyclic mapping, in which the adversarial loss is replaced with the classification adversarial loss and the defect adversarial loss to consolidate the forward mapping. Compared with conventional baseline methods, it can learn feature mapping between bone-conducted speech and target speech without additional air-conducted speech assistance. Moreover, the proposed method also avoids the over-smooth problem which is occurred commonly in conventional statistical based models. Experimental results show that the proposed method outperforms baseline methods such as CycleGAN, GMM, and BLSTM.
\keywords{Bone-conducted speech enhancement \and dual adversarial loss \and Parallel CycleGAN \and high frequency speech reconstruction}
\end{abstract}
]

{
\renewcommand{\thefootnote}{} 
\footnotetext{\Letter Jian Zhou}
\footnotetext{+86-152-5510-3835}
\footnotetext{jzhou@ahu.edu.cn}
\footnotetext{$^*$ Corresponding author}

\renewcommand{\thefootnote}{}
\footnotetext{}  

\renewcommand{\thefootnote}{}
\footnotetext{Qing Pan}
\footnotetext{e19301131@stu.ahu.edu.cn}

\renewcommand{\thefootnote}{}
\footnotetext{}

\renewcommand{\thefootnote}{}
\footnotetext{Teng Gao}
\footnotetext{e19301114@stu.ahu.edu.cn}

\renewcommand{\thefootnote}{}
\footnotetext{}

\renewcommand{\thefootnote}{}
\footnotetext{Huabin Wang}
\footnotetext{wanghuabin@ahu.edu.cn}

\renewcommand{\thefootnote}{}
\footnotetext{}

\renewcommand{\thefootnote}{}
\footnotetext{Liang Tao}
\footnotetext{taoliang@ahu.edu.cn}

\renewcommand{\thefootnote}{}
\footnotetext{}

\renewcommand{\thefootnote}{}
\footnotetext{Hon Keung Kwan}
\footnotetext{kwan1@uwindsor.ca}

\renewcommand{\thefootnote}{}
\footnotetext{}

\renewcommand{\thefootnote}1   
{
\fnsymbol{footnote}}
\footnotetext[1]{Key Laboratory of Intelligent Computing and Signal Processing of Ministry of Education, Anhui University, Hefei 230601, China}

\renewcommand{\thefootnote}2
{
\fnsymbol{footnote}}
\footnotetext[2]{Department of Electrical and Computer Engineering, University of Windsor, Windsor, ON N9B 3P4, Canada}

}

\section{Introduction}
\label{intro}
Speech is the most convenient way of communication and plays an important role for human-computer interaction. When a person is vocalizing, the vocal cord vibration signals are radiated through the acoustic cavity and the lips, producing air-conducted sound that are transmitted in the air. Meanwhile, this voicing behavior is also accompanied by skull and throat vibrations. The bone conduction microphone can collect such weak signals from the skull and the larynx to obtain the bone-conducted (BC) speech. 

Compared with the air-conducted (AC) speech, the BC speech is insusceptible of adverse environment noise due to its distinctive generation mechanism. The BC speech is suitable for communication in excessive noise environment~\cite{1} and has a wide range of applications in military and smart civilian fields~\cite{2}. However, because of the low-pass characteristics of skull, the voice signal collected by the bone conduction microphone loses many high frequency components (the spectrum components above 3kHz almost vanishes in Mandarin), resulting in that the BC speech sounds dull, and its intelligibility is low. 

To improve the speech quality and intelligibility of the BC speech, early methods usually adopt the source-filter model to decompose the BC speech into spectral envelope and excitation features, and mainly focus on the conversion of low-dimensional spectral envelope of the BC speech to that of the AC speech~\cite{1,3,4,6} because the human ear is more sensitive to spectral envelope~\cite{7}. In~\cite{8}, the Modulation Transfer Function (MTF) is used to decompose the BC speech into excitation signals and energy envelope in different frequency bands. In order to obtain a better decomposition and synthesis effect, a linear predictive coding (LPC) model is used instead of MTF in~\cite{9}.

Recently, neural networks have been widely used for nonlinear transformation between spectral envelope~\cite{10,11,12,13,a12}. In order to improve the quality and intelligibility of the BC speech, Changyan~\emph{et al}. proposed a BC speech enhancement method based on Bi-directional Long Short-Term Memory (BLSTM) type of recurrent neural networks and attention mechanism (AB-BLSTM), which has achieved good results~\cite{14}. Moreover, fusion features have also been studied in BC speech enhancement~\cite{a2}. However, BC speech enhancement fused with AC speech features is less effective in a strong noise environment. Hence, enhancing BC speech independently has important theoretical significance and practical application.

In recent years, the generative adversarial network (GAN) has attracted much attention because of its ability in simulating data distribution. GAN has been applied in speech processing tasks such as speech bandwidth expansion (BWE)~\cite{15} and voice conversion (VC). The aim of speech bandwidth expansion is to generate missing high-frequency components from a low-resolution speech signal. In \cite{a3}, a new BWE method which uses regularization method is proposed to conduct stable GAN training. The generator network is composed of a convolutional autoencoder with an embedded one-dimensional convolution kernel, which can estimate the high-frequency logarithmic power spectrum from the low-frequency logarithmic power spectrum. Kaneko \emph{et al}. applied GAN to sequence to sequence (Seq2Seq) VC~\cite{a4}. Experimental results show that GAN-based training is better than traditional mean-square error-based training.

The CycleGAN is a cycle-consistent adversarial networks with gated convolution and identity mapping loss, originally used for unpaired image style conversion~\cite{a1}. In a CycleGAN, the adversarial loss does not require explicit density estimation, which can reduce the negative impact of over-smoothing~\cite{a4,a5,a6,a7}. Recently, CycleGAN has been applied to solve VC problem~\cite{15}, by adding a gated convolution unit in the VC model and considering the identity mapping loss in the training phase to maintain the timing while ensuring that the semantic information unchanged.

In this paper, we propose a novel CycleGAN with dual adversarial loss (CycleGAN-DAL) model to conduct BC speech enhancement. The innovations of the proposed model are three-fold. First, the proposed model does not need to consider the relationship between the missing high-frequency components of the BC speech and the low-frequency components of the AC speech. Second, dual adversarial loss is suggested in the proposed model to train the discriminator in which the adversarial loss is divided into a classification adversarial loss and a defect adversarial loss. Third, the proposed model is in fact a blind enhancement method which does not require prior information such as the high frequency components of the BC speech and the low-frequency components of the AC speech. Demos of the present research can be viewed at ~\url{https://qpan77.github.io/Dadv_Cycle/demo.html}.

\section{CycleGAN-DAL Architectures for BC Speech Enhancement}
\label{sec:1}

\begin{figure}[b]
  \centering
  \includegraphics[scale=0.5]{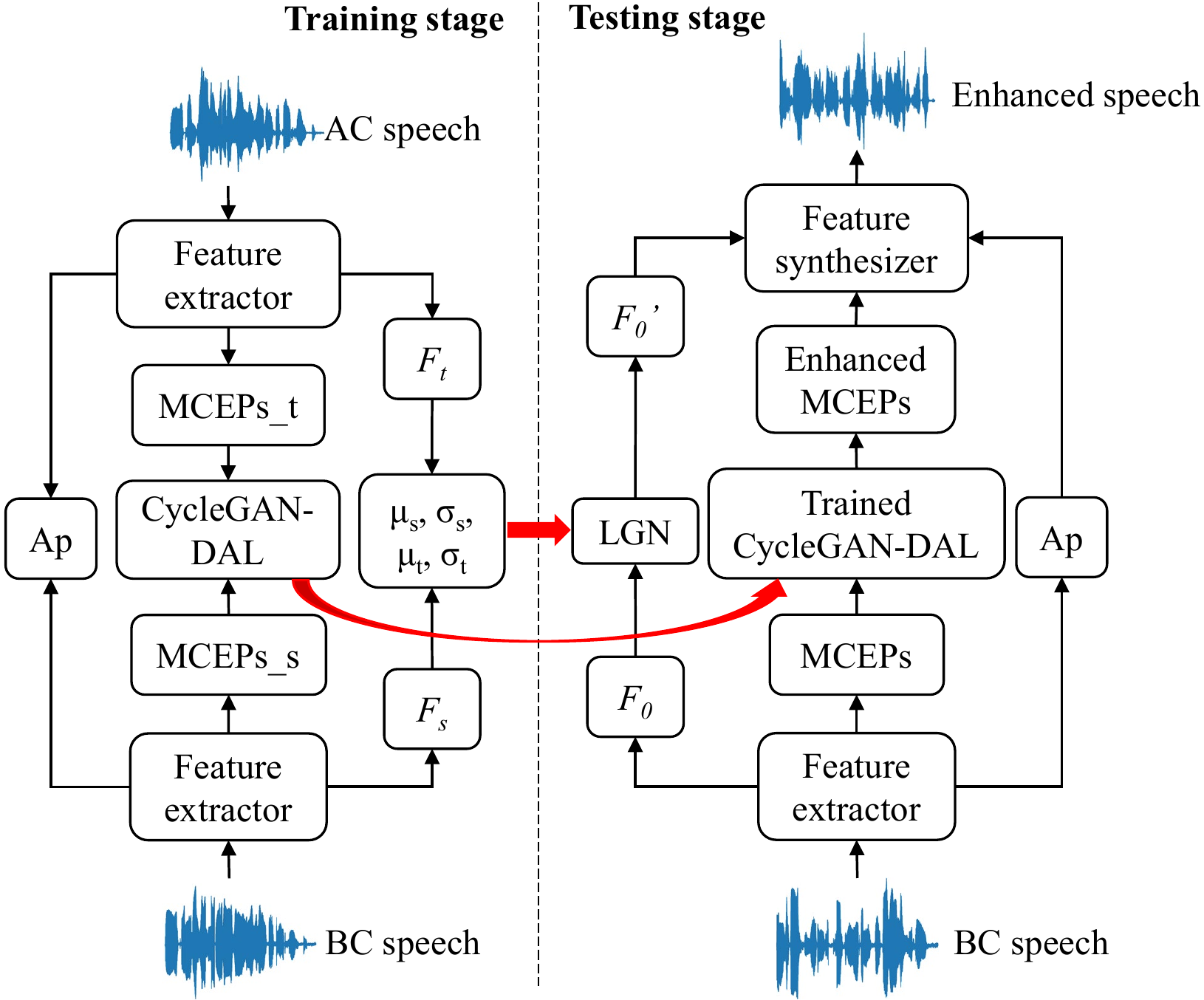}
  \caption{Architecture of the proposed method.} 
  \label{fig:1}
\end{figure}

\subsection{Architecture}
The architecture of the proposed method is depicted in Fig.~\ref{fig:1}. Firstly, the analysis model of the WORLD vocoder~\cite{a8} is applied to extract the feature parameters of the speech include spectral envelope (Sp), aperiodic parameter (Ap), and fundamental frequency ($F_0$). The Sp is further compressed to mel-cepstral coefficients (MCEPs). The MCEPs of the BC speech (MCEPs\_s) and the AC speech (MCEPs\_t) are used for training to learn the relationship between source features and target features. After training, the enhanced MCEPs features are obtained from the trained model. The conversion of $F_0$ from the BC speech to $F_0'$ of the enhanced speech is performed by the Logarithm Gaussian Normalization (LGN)~\cite{a9}. The enhanced $F_0'$ is calculated by 
\begin{equation}
log(F_0')=\frac{log(F_0)-\mu_{s}}{\sigma_{s}}\sigma_{t}+\mu_{t}
\label{con:log}
\end{equation}
where subscripts $s$ and $t$ represents BC speech and AC speech respectively.
Since existing studies~\cite{a10} shows that the Ap did not significantly affect the speech quality, the Ap of the enhanced speech is equal to that of the BC speech.

\subsection{Cycle-Consistent Loss and Identity Mapping Loss}
The forward mapping of the CycleGAN consists of two generators $G_{B \rightarrow A}$, $G_{A \rightarrow B}$ and one discriminator $D_A$~\cite{a1}. For BC enhancement tasks, let $\bm{b}\in R^{Q\times T_b}$ and $\bm{a}\in R^{Q\times T_a}$ represent respectively the features of the source BC speech $B$ and the features of the target AC speech $A$; $Q$ represents the dimension of the feature, $T_a$ and $T_b$ represents the dimension of the sequence length. For the CycleGAN shown in Fig.~\ref{fig:2}, the generator $G_{B \rightarrow A}$ maps the acoustic features $\bm{b}$ of the BC speech to the acoustic features $G_{B\rightarrow A}(\bm{b})$ of the predicted AC speech, and the generator $G_{A \rightarrow B}$ maps the converted predicted AC speech features $G_{B\rightarrow A}(\bm{b})$ to the predicted BC speech features $G_{A\rightarrow B}(G_{B\rightarrow A}(\bm{b}))$, and then the discriminator judges whether the features $G_{A\rightarrow B}(G_{B\rightarrow A}(\bm{b}))$ are consistent with the BC features $\bm{b}$ according to Eq.~(\ref{con:cyc}). The networks are iteratively trained for convergence.

The adversarial loss only judges whether $G_{B \rightarrow A}(\bm{b})$ obeys the target distribution and does not retain contextual information of $\bm{b}$, so the adversarial loss does not necessarily guarantee that the semantic information of $\bm{b}$ and $G_{B \rightarrow A}(\bm{b})$ are the same. To address this issue, we employ a cycle-consistent loss to keep the input and output features consistent in the aspect of semantics, the cycle-consistent loss~\cite{a12} is defined by
\begin{equation}
\begin{split}
L_{cyc}\left(G_{B \rightarrow A}, G_{A \rightarrow B}\right)=
E_{\bm{b} \sim P_{B}(\bm{b})}[\left\|G_{A \rightarrow B}\left(G_{B \rightarrow A}(\bm{b})\right)-\bm{b}\right\|_1]
\end{split}
\label{con:cyc}
\end{equation}
where $\|\cdot\|_1$ denotes $l_1$-norm, $P_{B}(\bm{b})$ denotes the distribution of BC speech data. $E_{\bm{b}\sim P_B (\bm{b})}$ denotes the expectations of the BC speech distribution. The cycle-consistent loss encourages $G_{B \rightarrow A}$ and $G_{A \rightarrow B}$ to find $(\bm{b}, \bm{a})$ pairs with the same contextual information.

Although the cycle-consistent loss provides constraints on the structure of spectral features, it is difficult to maintain the consistency of semantic information for a long sequence. We employ an identity mapping loss~\cite{a13} in the proposed model to consolidate contextual information, which is defined by
\begin{equation}
L_{id}(G_{B \rightarrow A}) = E_{\bm{a} \sim P_A(\bm{a})}[\left\|G_{B \rightarrow A}(\bm{a})-\bm{a} \right\|_1]
\label{con:id}
\end{equation}
where $P_A(\bm{a})$ denotes the distribution of AC speech data.

\subsection{CycleGAN Model for BC Speech Enhancement}

\begin{figure}[b]
  \centering
  \includegraphics[scale=0.4]{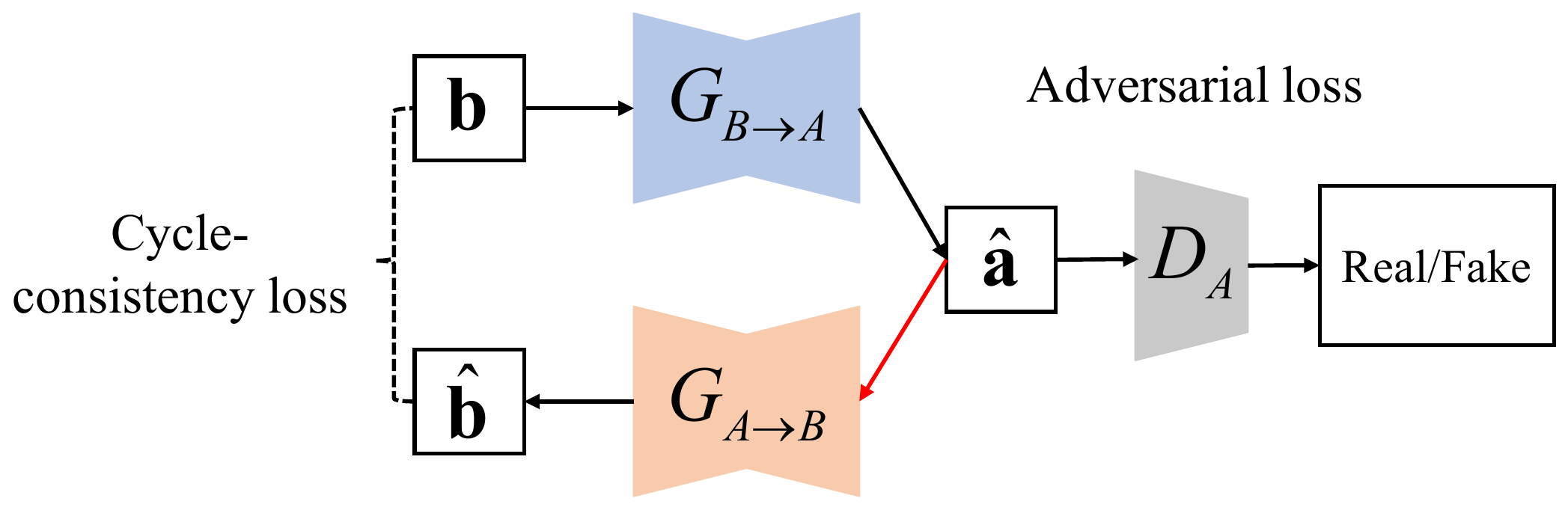}
  \caption{CycleGAN model.} 
  \label{fig:2}
\end{figure}

The original CycleGAN shown in Fig.~\ref{fig:2} adopts the adversarial loss \cite{16} to measure the degree of difference between the transformed features $G_{B \rightarrow A}(\bm{b})$ and the real features $\bm{a}$. The closer the transformed distribution $P_{G_{B \rightarrow A}(\bm{b})}$ to the real distribution $P_A(\bm{a})$, the smaller the loss function. The objective function is defined by
\begin{equation}
\begin{split}
L_{adv}\left(G_{B\rightarrow A}, D_A\right) = E_{\bm{a} \sim P_A(\bm{a})}[logD_A(\bm{a})] 
\\+ E_{\bm{b} \sim P_B(\bm{b})}[log(1-D_A(G_{B\rightarrow A}(\bm{b})))]
\end{split}
\label{con:adv}
\end{equation}
where $D_A(\bm{a})$ denotes the output of discriminator $D_A$ and $log$ refers to taking natural logarithm. $E_{\bm{a}\sim P_A (\bm{a})}$ denotes the expectations of the AC speech distribution. The generator $G_{B \rightarrow A}$ generates data that can deceive the discriminator $D_A$ by minimizing the loss, and the discriminator tries to be not deceived by the generator by maximizing the loss.

The objective loss function $L$ of the BC speech enhancement is defined by
\begin{equation}
\begin{split}
L=&L_{adv}\left(G_{B \rightarrow A}, D_A\right)+\lambda_{cyc} L_{cyc}\left(G_{B \rightarrow A}, G_{A \rightarrow B}\right)
\\&+\lambda_{id} L_{id}(G_{B \rightarrow A})
\end{split}
\end{equation}
where $\lambda_{cyc}$ and $\lambda_{id}$ denote respectively the trade-off coefficients of the loss functions $L_{cyc}\left(G_{B \rightarrow A}, G_{A \rightarrow B}\right)$ and $L_{id}(G_{B \rightarrow A})$.

\subsection{Proposed CycleGAN-DAL Model for BC Speech Enhancement}
\begin{figure}[b]
  \centering
  \includegraphics[scale=0.7]{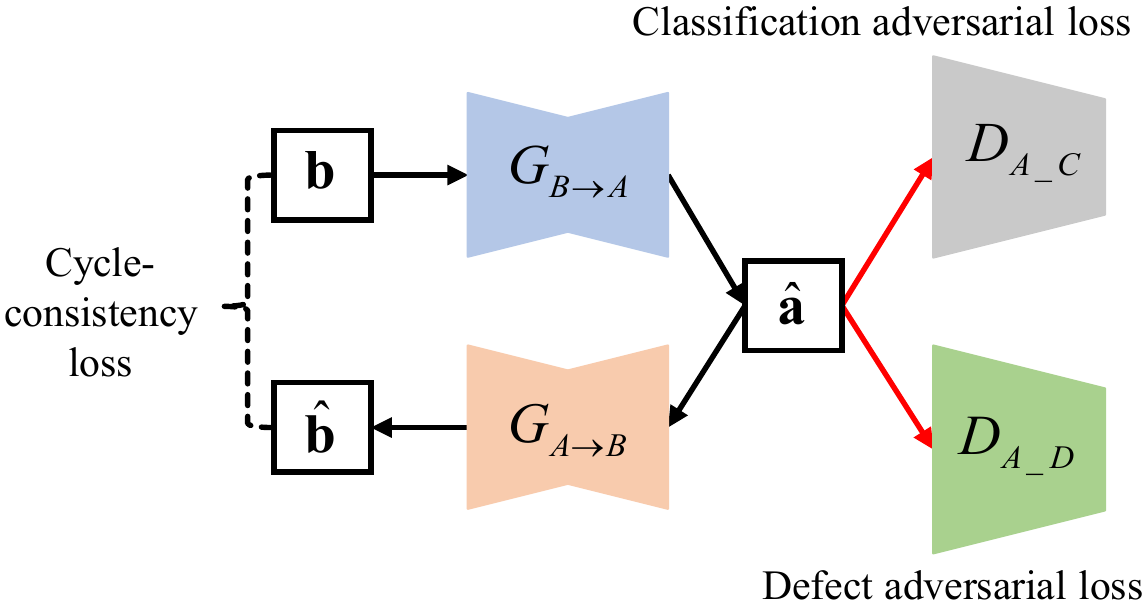}
  \caption{CycleGAN-DAL.} 
  \label{fig:3}
\end{figure}

To improve the effect of cyclic adversarial learning, we use two adversarial loss (consisting of a classification adversarial loss and a defect adversarial loss), the cycle-consistent loss and the identity mapping loss jointly control the mapping and learning process of CycleGAN-DAL. Both the BC speech and the AC speech are recorded simultaneously in one utterance, so their parallel corpus is obtained. Under normal circumstances, a parallel mapping is adopted for BC speech enhancement where the adversarial loss is replaced by two different adversarial losses, i.e., a classification adversarial loss and a defect adversarial loss as shown in Fig.~\ref{fig:3}. The classification adversarial loss function $L_{adc}\left(G_{B\rightarrow A}, D_{A_{-}C}\right)$ is responsible for classifying the generated speech features $G_{B \rightarrow A}(\bm{b})$ and the real speech features $\bm{a}$, which is defined by 
\begin{equation}
\begin{split}
L_{adc}\left(G_{B\rightarrow A}, D_{A_{-}C}\right) = E_{\bm{a} \sim P_A(\bm{a})}[logD_{A_{-}C}(\bm{a})] 
\\+ E_{\bm{b} \sim P_B(\bm{b})}[log(1-D_{A_{-}C}(G_{B\rightarrow A}(\bm{b})))]
\end{split}
\end{equation}
where $D_{A_{-}C}$ denotes the classification discriminator.

The defect adversarial loss function $L_{add}\left(G_{B\rightarrow A}, D_{A_{-}D}\right)$ is used to judge whether the expanded spectrum $G_{B \rightarrow A}(\bm{b})$ of the BC speech is damaged compare with the real AC speech spectrum $\bm{a}$, which is defined by
\begin{equation}
\begin{split}
L_{add}\left(G_{B\rightarrow A}, D_{A_{-}D}\right) = E_{a \sim P_A(a)}[logD_{A_{-}D}(a)] 
\\+ E_{b \sim P_B(b)}[log(1-D_{A_{-}D}(G_{B\rightarrow A}(b)))]
\end{split}
\label{con:add}
\end{equation}
where $D_{A_{-}D}$ denotes the defect discriminator.

Applying two different adversarial losses to replace the original adversarial loss provides dual constraints on category and defect is conducive to generate predicted features which are like real features to a greater extent.

The objective loss function $L_{dual}$ denotes the training target of the BC speech enhancement of the proposed method, which is defined by
\begin{equation}
\begin{split}
L_{dual}=&L_{adc}\left(G_{B \rightarrow A}, D_{A_{-}C}\right)+L_{add}\left(G_{B \rightarrow A}, D_{A_{-}D}\right)
\\&+\lambda_{cyc} L_{cyc}\left(G_{B \rightarrow A}, G_{A \rightarrow B}\right)
+\lambda_{id} L_{id}(G_{B \rightarrow A})
\end{split}
\end{equation}
The four loss functions are jointly used to control the stability of training for gradually approaching the real data.

Algorithm 1 describes the overall optimization procedure of CycleGAN-DAL. According to the algorithm, the generator and the discriminator are optimized respectively during the training process.

\begin{algorithm}
\caption{Training CycleGAN-DAL}
\begin{algorithmic}[1] 
\Require
BC speech feature: $B$; AC speech feature: $A$; Number of iterations: N
\Ensure
Trained model; Objective function $L_{dual}$

\For{n = 1 : N}
\State Sample $\bm{b} \in R^{Q\times T_b}$, $\bm{a} \in R^{Q\times T_a}$ from $B$, $A$

\State  $G_{B \rightarrow A}(\bm{b})$ $\leftarrow$ Generate enhanced features from $\bm{b}$ with $G_{B \rightarrow A}$

\State  $G_{A \rightarrow B}(G_{B \rightarrow A}(\bm{b}))$ $\leftarrow$ Generate cycle-consistent features from  
\State \ \ \ \ \ \ \ \ \ \ \ \ \ \ \ \ \ \ \ \ \ \ \ \ \ \ \ \ \ \ \ \ \ \  $G_{B \rightarrow A}(\bm{b})$ with $G_{A \rightarrow B}$

\State Compute cycle-consistent loss $L_{cyc}$

\State $G_{B \rightarrow A}(\bm{a})$ $\leftarrow$ Generate identity mapping features from $\bm{a}$ with 
\Statex \ \ \ \ \ \ \ \ \ \ \ \ \ \ \ \ \ \ \ \ \ \ \ \ \ \ \ \  $G_{B \rightarrow A}$
\State Compute identity mapping loss $L_{id}$

\State $D_{A\_C}(G_{B \rightarrow A}(\bm{b}))$ $\leftarrow$ Classify fake features with $D_{A\_C}$
\State Compute classification adversarial loss $L_{adc}$

\State $D_{A\_D}(G_{B \rightarrow A}(\bm{b}))$ $\leftarrow$ Measuring the similarity of fake spectra \Statex \ \ \ \ \ \ \ \ \ \ \ \ \ \ \ \ \ \ \ \ \ \ \ \ \ \ \ \ \ \ \ \ \ \ \ \ \ \   with $D_{A\_D}$
\State Compute defect adversarial loss $L_{add}$

\Statex \emph{Optimizing generator} :

\State $g$ $\leftarrow$ $\nabla(E_{\bm{b} \sim P_B(\bm{b})}[log(1-D_{A_{-}C}(G_{B\rightarrow A}(\bm{b})))]$
\Statex \ \ \ \ \ \ \ \ \ \ \ \ \ \ \ \  $+E_{b \sim P_B(\bm{b})}[log(1-D_{A_{-}D}(G_{B\rightarrow A}(\bm{b})))]+L_{cyc}+L_{id})$

\Statex \emph{Optimizing discriminator} :
\State $g$ $\leftarrow$ $\nabla(E_{\bm{a} \sim P_A(\bm{a})}[logD_{A_{-}C}(\bm{a})]+E_{a \sim P_A(a)}[logD_{A_{-}D}(a)])$

\State $L_{dual}$ $\leftarrow$ Objective function to be minimized
\EndFor
\end{algorithmic}
\end{algorithm}

\begin{figure}
  \centering
  \includegraphics[scale=0.5]{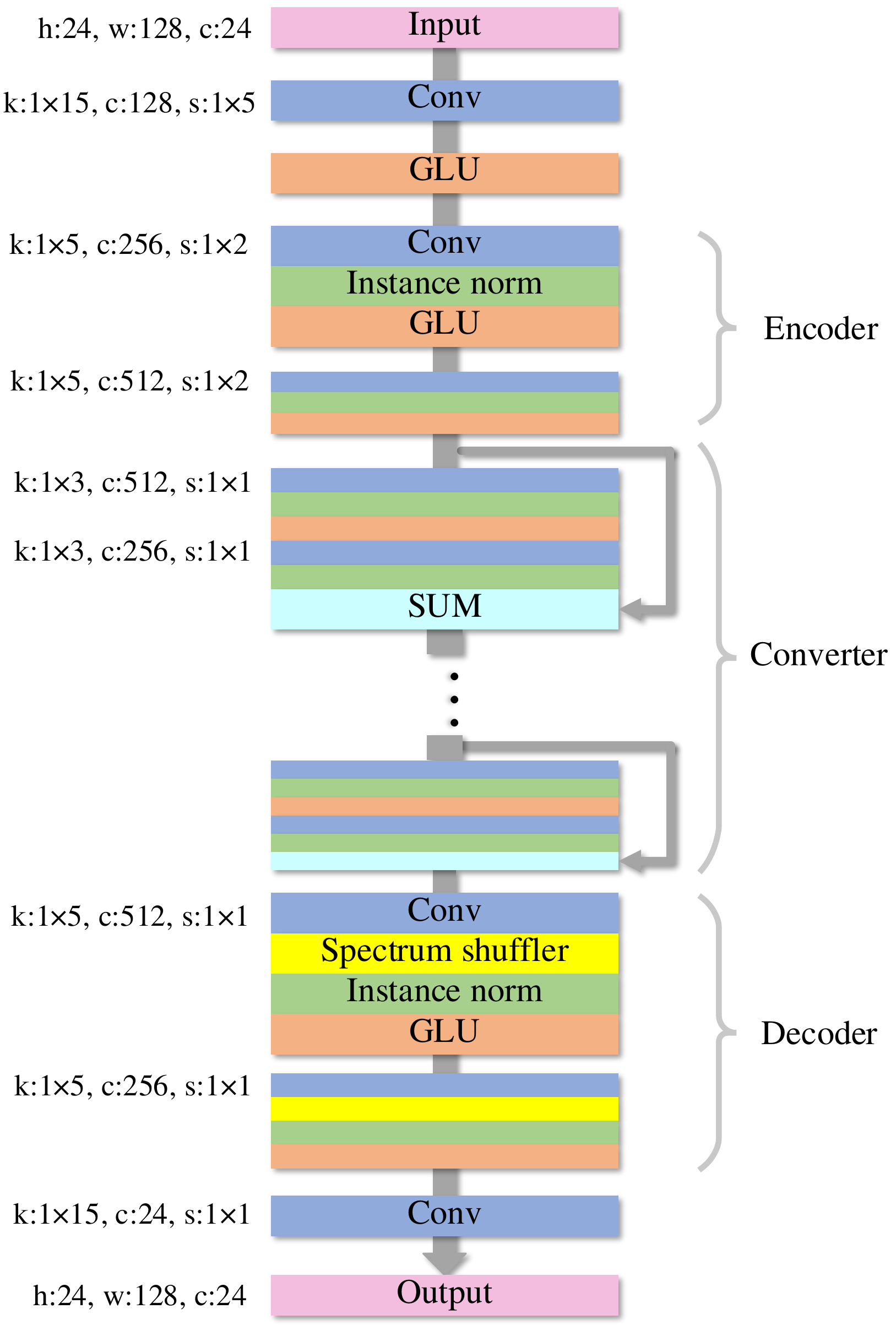}
  \caption{Generator of CycleGAN-DAL.} 
  \label{fig:4}
\end{figure}

\subsection{CycleGAN and CycleGAN-DAL Structure}
\subsubsection{Generator}
The generator of the proposed CycleGAN-DAL is shown in Fig.~\ref{fig:4}. For comparison, the same generator is also used in the CycleGAN. The generator is composed of three parts: Encoder, converter and decoder. The encoder contains two one-dimensional convolutional network blocks, each of which is composed of a gated convolutional layer, an instance normalization (IN) layer and an activation function layer. The convolution layer uses a convolution kernel with a size of 1×5, 1 represents the time dimension and 5 corresponds to the feature dimension of the MCEPs. The encoder captures the feature and retains the time structure, maps the feature to a lower-dimensional space, and expands the bandwidth in parallel. The IN layer calculates the mean and standard deviation of all the spectral values in each feature map and then normalizes it to speed up the convergence of the model and maintain the independence between features in each frame. Like text, speech has a sequential structure, and the current output is related to the previous output. But convolutional networks can be parallelized without this dependence. The activation function of the method in this paper uses Gated Linear Units (GLUs)~\cite{a11} to achieve speech modeling,
\begin{equation}
O_{x+1}=(O_x\cdot W_x+b_x)\otimes \sigma (O_x\cdot V_x+c_x)
\end{equation}
where $W_x$ and $V_x$ represent the weights of the $x$ layer, $b_x$ and $c_x$ represent the biases of the $x$ layer, $O_{x+1}$ is the output of ${x+1}$ layer, $\otimes$ is the element product, $\sigma$ is the activation function. The GLU can selectively transmit information according to the state of the previous layer.


The converter consists of six residual network blocks, each of which contains the residual connection and the skip connection. The residual network structure helps to train the deep network and to accelerate the convergence of the network. The residual network is expressed as,
\begin{equation}
R_{l+1}=H_{l+1}(R_l)+R_l
\end{equation}
where $R_l$ and $R_{l+1}$ are the input and the output of the ${l+1}$ residual block, $H_{l+1}$ indicates two blocks and each block consists of convolution, IN and ReLU activation function.

The decoder is composed of a deconvolution network, the basic structure of which is the same as that of the encoder. In addition, a spectrum shuffling layer is added in front of the convolution layer to improve speech intelligibility of the enhanced speech. The decoding process can be regarded as upsampling of the predicted feature to a high-dimensional feature to obtain converted speech features. In Fig. ~\ref{fig:4}, $h$, $w$, and $c$ represent the height, width, and number of channels, respectively, and $k$, $c$, and $s$ represent the size of the convolution kernel, the number of channels, and the stride, respectively.

\subsubsection{Discriminator}
\begin{figure}[b]
  \centering
  \includegraphics[scale=0.5]{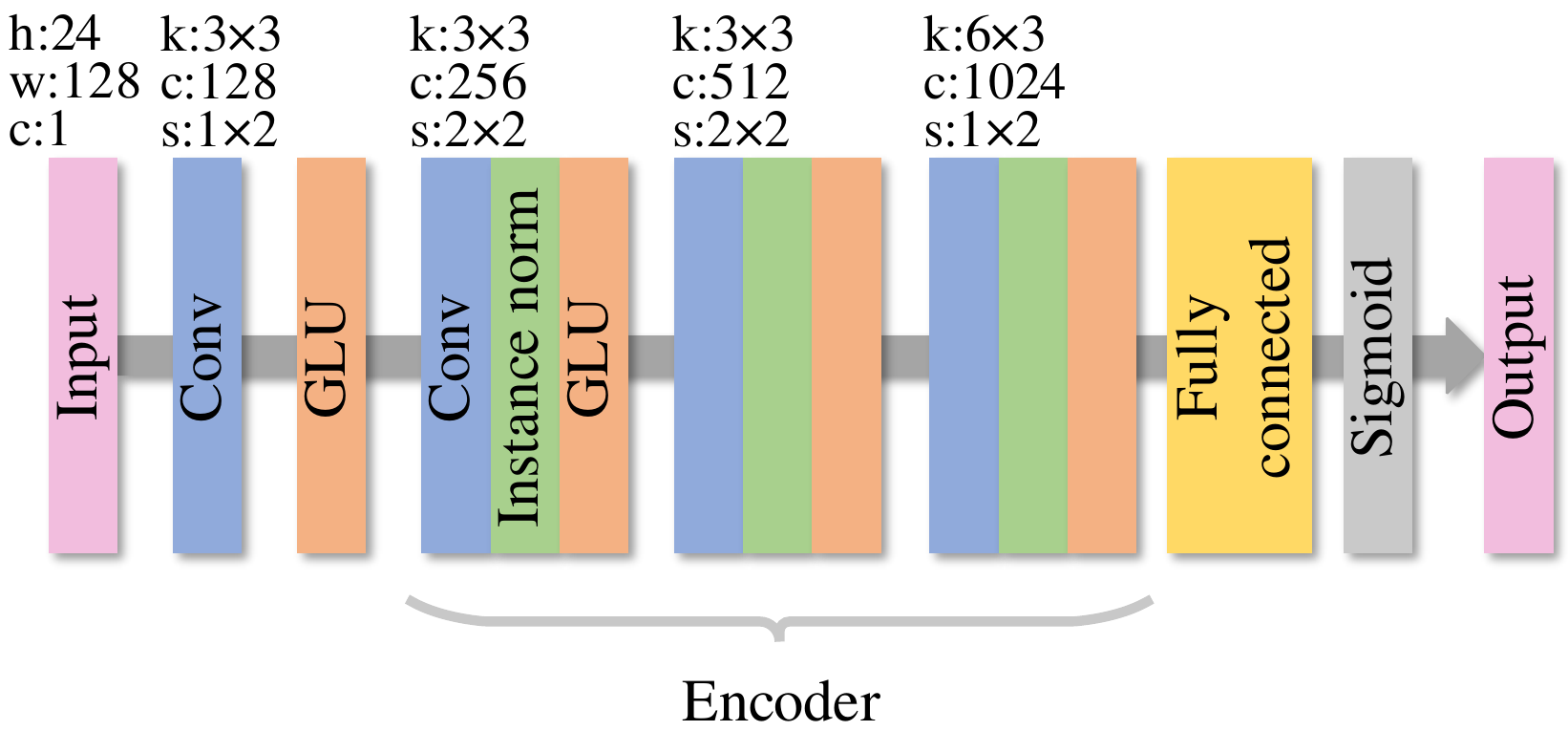}
  \caption{Discriminator of CycleGAN-DAL.} 
  \label{fig:5}
\end{figure}

The discriminator used in the proposed CycleGAN-DAL is shown in Fig.~\ref{fig:5}. The same discriminator is also used in the CycleGAN for comparison. The discriminator uses a two-dimensional convolutional network to distinguish between spectral feature generated by the generator and real spectral feature, aiming to encourage the generator to model the distribution of the target data and to avoid over-smoothing. To this end, the discriminator first smooths one-dimensional speech feature into two-dimensional spectral feature. Then the input features are encoded, discriminator uses a fully connected layer to determine the authenticity of the overall structure, and then maps the samples to the labels. Finally, the sigmoid activation function obtains a value representing the degree of trust, which is used for calculating the loss function.

\section{Experiments}
\subsection{Configuration}
To verify the effectiveness of the proposed CycleGAN-DAL model, we perform BC speech enhancement on two public BC speech dataset, i.e., the dataset proposed in~\cite{17} (denoted as dataset A) and the dataset proposed in~\cite{18} (denoted as dataset B). The dataset A comprises 200 sentences, each of which is uttered by two native speakers respectively. The dataset B is a Mandarin corpus dataset which comprises 320 sentences, each of which is uttered by 6 speakers. All BC speech is downsampled to 16 kHz. The WORLD vocoder is adopted to extract 24-dimensional MCEPs, $F_0$ and Ap every 5 $ms$. In the experiment, the training set and the test set are divided according to a ratio of 4:1

The parameter configuration of the generator and the discriminator in the proposed CycleGAN-DAL model are listed in Table~\ref{tab:table1}. To improve the model stability, least squares loss is adopted instead of negative log-likelihood target adversarial loss. The Adam optimizer is used and the trade-off coefficient $\lambda_{cyc}$ is set to 10, with $\lambda_{id}$ to 5, and the batch to 1. The learning rate of the generator is 0.0002, while the learning rate of the discriminator is 0.0001. The training epoch is set to 3000. 

Two feature mapping strategies are evaluated. For the nonparallel mapping strategy, the source BC features and the target AC features are randomly selected as model input. For the parallel mapping strategy, the source BC features and their corresponded AC target features are selected as model input.
  
\newcommand{\tabincell}[2]{\begin{tabular}{@{}#1@{}}#2\end{tabular}}

\begin{table}[]
\caption{Parameter settings of the generator and the discriminator of CycleGAN-DAL}
\label{tab:table1}
\resizebox{8.5cm}{!}
{
\centering
\begin{tabular}{@{}lllll@{}}
\midrule
                               & Module                          & Kernel size          & Stride size         & Channels         \\ 
\toprule
\multirow{9}{*}{Generator}     & Conv                            & 1×15                 & 1×1                 & 128              \\ \cmidrule(l){2-5} 
                               & \multirow{2}{*}{Encoder}     & 1×5                  & 1×2                 & 256              \\
                               &                                 & 1×5                  & 1×2                 & 512              \\ \cmidrule(l){2-5} 
                               & \multirow{3}{*}{Converter} & \multicolumn{3}{l}{6 residual blocks have the same structure} \\
                               &                                 & 1×3                  & 1×1                 & 1024             \\
                               &                                 & 1×3                  & 1×1                 & 512              \\ \cmidrule(l){2-5} 
                               & \multirow{2}{*}{Decoder}       & 1×5                  & 1×1                 & 1024             \\
                               &                                 & 1×5                  & 1×1                 & 512              \\ \cmidrule(l){2-5} 
                               & Conv                            & 1×15                 & 1×1                 & 24               \\ \midrule
\multirow{5}{*}{Discriminator} & Conv                            & 3×3                  & 1×2                 & 128              \\ \cmidrule(l){2-5} 
                               & \multirow{3}{*}{Encoder}     & 3×3                  & 2×2                 & 256              \\
                               &                                 & 3×3                  & 2×2                 & 512              \\
                               &                                 & 6×3                  & 1×2                 & 1024             \\ \cmidrule(l){2-5} 
                               &                               & \multicolumn{3}{l}{Fully connected layer}      \\ 
\hline
\end{tabular}
}
\end{table}
 
\subsection{Results and Disscussion}
In order to show the effectiveness of the proposed CycleGAN-DAL model, multiple sets of comparative experiments were carried out. The Gaussian mixture model (GMM) and the BLSTM are trained respectively to conduct BC speech enhancement. The GMM is divided into maximum likelihood estimation GMM\_w and non-maximum likelihood estimation GMM\_wo.  In addition, both the CycleGAN model with nonparallel mapping strategy (denoted as NMC) and the CycleGAN model with parallel mapping strategy (denoted as PMC) are used to conduct BC speech enhancement. For the proposed CycleGAN-DAL model, we only consider parallel mapping strategy for BC spech enhancement (denoted as PMCD).

\begin{figure*}[htbp]
  \centering
  \includegraphics[width=1\textwidth]{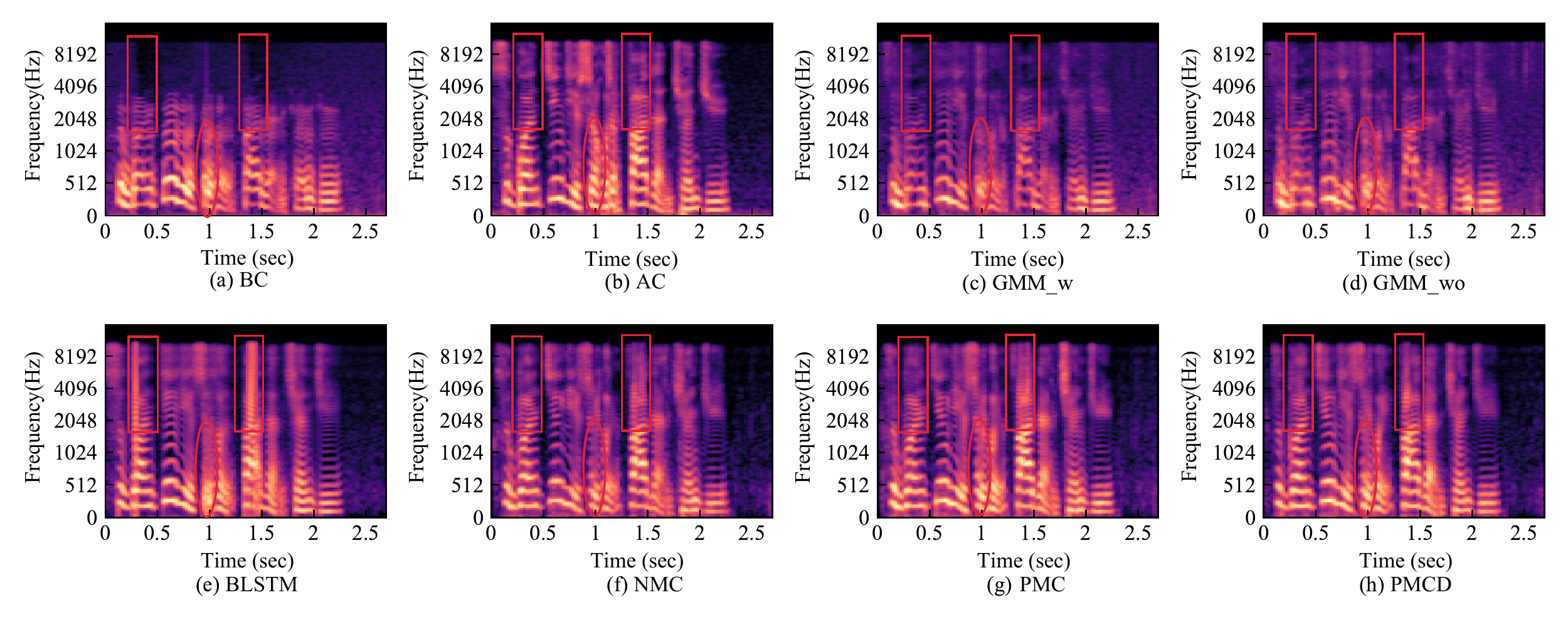}
  \caption{Enhanced speech spectrums obtained by different methods.} 
  \label{fig:6}
\end{figure*}

The spectrums of the enhanced speech obtained by different methods are shown in Fig.~\ref{fig:6}, it can be seen that the PMCD method obtains the spectrum most similar to the AC speech  at high frequency as shown in the red rectangular-shape box, while the two GMM methods have poor recovering ability at high frequency and the energies of the spectrums obtained by the two GMM methods are low. The resolution of the spectrum obtained by the BLSTM method are also relatively low. In Fig.~\ref{fig:6}, the low frequency components in the red elliptical-shape box show that methods based on CycleGAN get a harmonic structure closer to the AC speech.

To evaluate the quality and intelligibility of the enhanced speech, short-time objective intelligibility (STOI) ~\cite{19}, log-spectral distance (LSD)~\cite{20}, and P.563~\cite{21} are used for objective evaluation. STOI is a measure of speech intelligibility, ranging from 0 to 1. The higher the score, the higher the speech intelligibility after enhancement. STOI is defined by
\begin{equation}
d = \frac{1}{JM}\sum_{j,m}d_j(m)
\end{equation}
where
\begin{equation}
X_j(m)=\sqrt{\sum_{k=k_1(j)}^{k_2(j)-1}|\hat{x}(k, m)|^{2}}
\end{equation}
\begin{equation}
Y'=max(min(\alpha Y, X+10^{-\beta /20}X), X-10^{-\beta /20}X)
\end{equation}
\begin{equation}
d_j(m) = \frac{\sum_n(X_j(n)-\frac{1}{N}\sum_lX_j(l))(Y_j^{'}(n)-\frac{1}{N}\sum_lY_j^{'}(l))}{\sqrt{\sum_n(X_j(n)-\frac{1}{N}\sum_lX_j(l))^{2}(Y_j^{'}(n)-\frac{1}{N}\sum_lY_j^{'}(l))^{2}}'}
\end{equation}
where $\hat{x}(k,m)$ represents the $k^{th}$ DFT-bin of the $m^{th}$ frame of the AC speech. The norm of the $j^{th}$ one-third octave band referred to as a TF-unit. $k_1$ and $k_2$ denote the one-third octave band edges, which are round to the nearest DFT-bin. The TF-representation of the enhanced speech is obtained similarly represented by $Y_j(m)$. The intermediate intelligibility measure for one TF-unit denoted by $d_j(m)$, depends on a region of $N$ consecutive TF-units from both $X_j(n)$ and $Y_j(n)$, where $n\in \mathcal{M}$ and $\mathcal{M}=\{(m-N+1), (m-N+2), … , m-1, m\}(N=30)$. A local normalization procedure is applied, by scaling all the TF-units from $Y_j(n)$ with a factor $\alpha=(\sum_nX_j(n)^{2}/\sum_nY_j(n)^{2})^{1/2}$, such that its energy equals the AC speech energy, within that TF-region. $Y'$ denotes the normalized and clipped TF-unit and $\beta(=-15)$ represents the lower signal-to-distortion ratio (SDR) bound. The $d_j(m)$ is defined as an estimate of the linear correlation coefficient between the AC and enhanced TF-units, where $l\in \mathcal{M}$. Finally, STOI is  given by the average of $d_j(m)$ over all bands and frames,where $M$ and $J$ denote the total number of frames and the number of one-third octave bands.

LSD measures the difference between the power spectrums of the enhanced speech $p(w)$ and the target speech $\hat{p}(w)$. The smaller the value, the smaller the distortion of the speech spectrum. The logarithmic spectral distance between $p(w)$ and $\hat{p}(w)$ is defined as
\begin{equation}
D_{L}=\sqrt{\frac{1}{2 \pi} \int_{-\pi}^{\pi}\left[10^{*} \log _{10} \frac{p(w)}{\hat{p}(w)}\right]^{2} d w}
\end{equation}

P.563 is a single-ended voice quality evaluation standard proposed by the ITU-T, which does not require the original signal and can directly output the fluency score of the signal~\cite{22}. Compared with the reference-based quality evaluation method, its usability is higher. The measurement procedure includes three stages: the preprocessing stage, the distortion estimation stage, and the perceptual mapping stage. The rank of the annoyance or perceptual focus was found by analyzing auditory experiments and is listed in Table~\ref{tab:table5}.

\begin{table}[]
  \caption{P.563 distortion classes}
  \label{tab:table5}
  \centering
  \begin{tabular}{@{}lll@{}}
   \midrule
  \makecell[c]{Order}  & \makecell[c]{Class discription} \\   \toprule
   \makecell[c]{1}   & High level backgroud noise \\
   \makecell[c]{2}   & Signal interruptions \\
   \makecell[c]{3}   & Signal correlated noise \\
   \makecell[c]{4}   & Speech robotization \\
   \makecell[c]{5}   & Common unnaturalness\\  \midrule
  \end{tabular}
 \end{table}


From Tables~\ref{tab:table2}, \ref{tab:table3}, and \ref{tab:table4}, the average performance of the PMCD method is better than the GMM and BLSTM methods, and is slightly better than the NMC and PMC methods. In general, the NMC and PMC methods are better than the GMM and BLSTM methods except the BLSTM method is marginally better than the NMC and PMC methods for the STOI and P.563 results using Dataset A. Because BLSTM is based on sequence modeling and considers the relationship between frames,  CycleGAN is parallelized and does not need to consider the relationship between frames, the subjective listening experience based on the CycleGAN is better.

\begin{table}[b]
 \caption{STOI evaluation of the enhanced speeches obtained by different BC speech enhancement methods.}
  \label{tab:table2}
  \resizebox{8.5cm}{!}
{
  \centering
\begin{tabular}{@{}lllllllll@{}}
\midrule
Corpus  & Speaker & BC   & \tabincell{c}{GMM\\\_w} & \tabincell{c}{GMM\\\_wo} & BLSTM         
&NMC & PMC & PMCD    \\ \toprule
\multirow{3}{*}{\tabincell{c}{Dataset\\A~\cite{17}}} & Female  & 0.63 & 0.68   & 0.65    & \textbf{0.80} & 0.78    & 0.78 & 0.79          \\
                           & Male    & 0.61 & 0.66   & 0.64    & 0.78          & 0.76    & 0.77 & \textbf{0.81} \\
                           &Average  &0.620 &0.670 &0.645 &0.790 &0.770 &0.775 &\textbf{0.800} \\
 \midrule
\multirow{7}{*}{\tabincell{c}{Dataset\\B~\cite{18}}} & 01      & 0.73 & 0.76   & 0.75    & \textbf{0.86} & 0.84    & 0.84 & 0.85          \\
                           & 02      & 0.70 & 0.71   & 0.69    & 0.80          & 0.82    & 0.83 & \textbf{0.84} \\
                           & 03      & 0.73 & 0.78   & 0.76    & \textbf{0.90} & 0.88    & 0.88 & 0.88          \\
                           & 04      & 0.68 & 0.73   & 0.71    & 0.84          & 0.80    & 0.80 & \textbf{0.85} \\
                           & 05      & 0.60 & 0.64   & 0.61    & 0.70          & 0.74    & 0.74 & \textbf{0.77} \\
                           & 06      & 0.68 & 0.74   & 0.72    & 0.80          & 0.84    & 0.85 & \textbf{0.86} \\
                           &Average &0.687 &0.727 &0.707 &0.817 &0.820  &0.823 & \textbf{0.842} \\
                            \midrule
\end{tabular}
}
\end{table}
 
\begin{table}[]
  \caption{LSD evaluation of the enhanced speeches obtained by different BC speech enhancement methods.}
  \label{tab:table3}
  \resizebox{8.5cm}{!}
  {
  \centering
  \begin{tabular}{@{}lllllllll@{}}
   \midrule
  Corpus  & Speaker  & BC & \tabincell{c}{GMM\\ \_w}  & \tabincell{c}{GMM\\ \_wo}  & BLSTM &NMC  &PMC  &PMCD  \\ \toprule
\multirow{3}{*}{\tabincell{c}{Dataset\\A~\cite{17}}}   & Female  & 1.73 & 1.45 & 1.41 & 1.08 &1.00 &1.00 &\textbf{0.99} \\
    & Male    & 1.82 & 1.48 & 1.44 & 1.16 &1.04 &1.03 &\textbf{1.01} \\
    &Average &1.775 &1.465 &1.425 &1.120 &1.020 &1.015 &\textbf{1.000} \\ \midrule
\multirow{7}{*}{\tabincell{c}{Dataset\\B~\cite{18}}}  & 01  & 1.58 & 1.32 & 1.31 & 1.07 &0.97 &0.98 &\textbf{0.96} \\
   & 02   & 1.21 & 1.00 & 1.00 & 0.98 &\textbf{0.93} &0.94 &0.94 \\
     & 03  & 1.19 & 1.06 & 1.05 & \textbf{0.89} &0.91 &0.90 &0.91 \\
     & 04 & 1.20 & 1.06 & 1.05 & 0.98 &0.99 &0.97 &\textbf{0.95} \\
   & 05   & 1.17 & 0.99 & 1.01 & 1.11 & 1.00 &1.00 &\textbf{0.99}\\
   & 06   & 1.06 & 0.97 & 0.96 & 0.92 &0.93 &0.94 &\textbf{0.92} \\ 
   &Average &1.235 &1.067 &1.063 &0.992 &0.955 &0.955 &\textbf{0.945} \\
   \midrule
  \end{tabular}
  }
 \end{table}

 In this paper, we adopt parallel mapping between corpus because of the advantage of synchronous collection of BC speech and AC speech. If there is a lack of parallel corpus, nonparallel mapping can also be established in practical situation.

 \begin{table}[]
  \caption{P.563 evaluation of the enhanced speeches obtained by different BC speech enhancement methods.}
  \label{tab:table4}
  \resizebox{8.5cm}{!}
  {
  \centering
  \begin{tabular}{@{}lllllllll@{}}
   \midrule
  Corpus  & Speaker & GMM\_w  & GMM\_wo  & BLSTM  &NMC  &PMC  &PMCD  \\   \toprule
   \multirow{3}{*}{\tabincell{c}{Dataset\\A~\cite{17}}}   & Female  & 3.05 & 2.88 & 4.85 & 4.86 &4.82 &\textbf{4.89} \\
   & Male    & 2.82 & 2.75 & 4.36 & 4.32 &\textbf{4.38} &4.35 \\   
   &Average &2.935 &2.815 &4.605 &4.590 &4.600 &\textbf{4.620} \\
   \midrule
  \multirow{7}{*}{\tabincell{c}{Dataset\\B~\cite{18}}} & 01  & 2.07 & 2.13 & 3.73 & 4.20 &4.26 &\textbf{4.31} \\
   & 02   & 2.66 & 2.72 & 4.79 & 4.93 &\textbf{4.94} &4.92 \\
      & 03  & 2.30 & 2.38 & 4.94 & 4.99 &4.97 &\textbf{5.00} \\
     & 04 & 2.44 & 2.46 & 4.90 & 4.91 &4.85 &\textbf{4.92} \\
   & 05   & 3.14 & 3.16 & 4.94 & 4.97 & 4.97 &\textbf{4.99}\\
   & 06   & 1.37 & 1.39 & 3.72 & 3.98 &4.08 &\textbf{4.15} \\
   &Average &2.330 &2.373 &4.503 &4.663 &4.678 &\textbf{4.715} \\  \midrule
  \end{tabular}
  }
 \end{table}

\section{Conclusion}
In this paper, we has proposed a parallel CycleGAN BC speech enhancement method with dual adversarial losses. This method replaces the adversarial loss with a classification adversarial loss and a defect adversarial loss. The classification adversarial loss distinguishes the generated speech from the real speech, and the defect adversarial loss measures the difference between the expanded spectrum of BC speech and the spectrum of AC speech. Experimental results show that the method can recover the missing high-frequency components and obtain spectrum with better harmonic structure, greatly improving the quality and intelligibility of BC speech, which is beneficial to ameliorating the quality of voice communication in adverse noise environment.

\section{Compliance with Ethical Standards}
\subsection{Conflict of Interest}
The authors declare that they have no conflict of interest.

\subsection{Ethical Approval}
This article does not contain any studies with human participants or animals performed by any of the authors.




\end{document}